# Testing Closeness of Discrete Distributions[*]

Tuğkan Batu[†]  Lance Fortnow[‡]  Ronitt Rubinfeld[§]  Warren D. Smith[¶]
Patrick White

May 29, 2018

**Abstract**

Given samples from two distributions over an $n$-element set, we wish to test whether these distributions are statistically close. We present an algorithm which uses sublinear in $n$, specifically, $O(n^{2/3}\epsilon^{-8/3}\log n)$, independent samples from each distribution, runs in time linear in the sample size, makes no assumptions about the structure of the distributions, and distinguishes the cases when the distance between the distributions is small (less than $\max\{\epsilon^{4/3}n^{-1/3}/32, \epsilon n^{-1/2}/4\}$) or large (more than $\epsilon$) in $\ell_1$ distance. This result can be compared to the lower bound of $\Omega(n^{2/3}\epsilon^{-2/3})$ for this problem given by Valiant [54].

Our algorithm has applications to the problem of testing whether a given Markov process is rapidly mixing. We present sublinear algorithms for several variants of this problem as well.

---

# 1 Introduction

Suppose we have two distributions over the same $n$-element set, such that we know nothing about their structure and the only access we have to these distributions is the ability to take independent samples from them. Suppose further that we want to know whether these two distributions are close to each other in $\ell_1$ norm.[1] A first approach, which we refer to as the *naive approach*, would be to sample enough elements from each distribution so that we can approximate the distribution and then compare the approximations. It is easy to see (see Theorem 23 in Section 3.2) that this naive approach requires the number of samples to be at least linear in $n$.

In this paper, we develop a method of testing that the distance between two distributions is at most $\epsilon$ using considerably fewer samples. If the distributions have $\ell_1$ distance at most $\max\{\epsilon^{4/3}n^{-1/3}/32, \epsilon n^{-1/2}/4\}$, then the algorithm will accept with probability at least $1 - \delta$. If the distributions have $\ell_1$ distance more than $\epsilon$ then the algorithm will accept with probability at most $\delta$. The number of samples used is $O(n^{2/3}\epsilon^{-8/3}\log(n/\delta))$. In contrast, the methods of Valiant [54], fixing the incomplete arguments in the original conference paper (see Section 3), yield an $\Omega(n^{2/3}\epsilon^{-2/3})$ lower bound for testing $\ell_1$ distance in this model.

Our test relies on a test for whether two distributions have small $\ell_2$ distance, which is considerably easier to test: we give an algorithm with sample complexity independent of $n$. However, small $\ell_2$ distance does not in general give a good measure of the closeness of two distributions according to $\ell_1$ distance. For example, two distributions can have disjoint support and still have $\ell_2$ distance of $O(1/\sqrt{n})$. Still, we can get a very good estimate of the $\ell_2$ distance, say to within $O(1/\sqrt{n})$ additive error, and then use the fact that the $\ell_1$ distance is at most $\sqrt{n}$ times the $\ell_2$ distance. Unfortunately, the number of queries required by this approach is too large in general. Because of this, our $\ell_1$ test is forced to distinguish between two cases.

For distributions with small $\ell_2$ norm, we show how to use the $\ell_2$ distance to get an efficient test for $\ell_1$ distance. For distributions with larger $\ell_2$ norm, we use the fact that such distributions must have elements which occur with relatively high probability. We create a filtering test that partitions the domain into those elements with relatively high probability and all the other elements (those with relatively low probability). The test estimates the $\ell_1$ distance due to these high-probability elements directly, using the naive approach mentioned above. The test then approximates the $\ell_1$ distance due to the low-probability elements using the test for $\ell_2$ distance. Optimizing the notion of "high probability" yields our $O(n^{2/3}\epsilon^{-8/3}\log(n/\delta))$ algorithm. The $\ell_2$ distance test uses $O(\epsilon^{-4}\log(1/\delta))$ samples.

Applying our techniques to Markov chains, we use the above algorithm as a basis for constructing tests for determining whether a Markov chain is rapidly mixing. We show how to test whether iterating a Markov chain for $t$ steps causes it to reach a distribution close to the stationary distribution. Our testing algorithm works by following $\tilde{O}(tn^{5/3})$ edges in the chain. When the Markov chain is dense enough and represented in a convenient way (such a representation can be computed in linear time and we give an example representation in Section 4), this test remains sublinear in the size of the Markov chain for small $t$. We then investigate two notions of being *close* to a rapidly mixing Markov chain that fall within the framework of property testing, and show how to test that a given Markov chain is close to a Markov chain that mixes in $t$ steps by following only $\tilde{O}(tn^{2/3})$ edges. In the case of Markov chains that come from directed graphs and pass our test, our theorems show the existence of a directed graph that is both close to the original one and rapidly mixing.

---

[1]Half of $\ell_1$ distance between two distributions is also referred to as total variation distance.

## 1.1 Related Work

**Testing Properties of Distributions**  The use of collision statistics in a sample has been proposed as a technique to test whether a distribution is uniform (see, for example, Knuth [40]). Goldreich and Ron [32] give the first formal analysis that using $O(\sqrt{n})$ samples to estimate the collision probability yields an algorithm which gives a very good estimate of the $\ell_2$ distance between the given distribution and the uniform distribution. Their "collision count" idea underlies the present paper. More recently, Paninski [45] presents a test to determine whether a distribution is far from the uniform distribution with respect to $\ell_1$ distance using $\Theta(\sqrt{n}/\epsilon^2)$ samples. Ma [42] also uses collisions to measure the entropy of a distribution defined by particle trajectories. After the publication of the preliminary version of this paper, a long line of publications appeared regarding testing properties of distributions including independence, entropy, and monotonicity (see, for example, [8, 10, 7, 12, 5, 54, 49, 48, 51, 1]).

**Expansion, Rapid Mixing, and Conductance**  Goldreich and Ron [32] present a test that they conjecture can be used to give an algorithm with $O(\sqrt{n})$ query complexity which tests whether a regular graph is close to being an expander, where by close they mean that by changing a small fraction of the edges they can turn it into an expander. Their test is based on picking a random node and testing whether random walks from this node reach a distribution that is close to the uniform distribution on the nodes. Our tests for Markov chains are based on similar principles. Mixing and expansion are known to be related [53], but our techniques only apply to the mixing properties of random walks on directed graphs, since the notion of closeness we use does not preserve the symmetry of the adjacency matrix. More recently, a series of papers [22, 37, 43] answer Goldreich and Ron's conjecture in the affirmative. In a previous work, Goldreich and Ron [31] show that testing that a graph is close to an expander requires $\Omega(n^{1/2})$ queries.

The conductance [53] of a graph is known to be closely related to expansion and rapid-mixing properties of the graph [38, 53]. Frieze and Kannan [27] show, given a graph $G$ with $n$ vertices and $\alpha$, one can approximate the conductance of $G$ to within additive error $\alpha$ in time $n \cdot 2^{\tilde{O}(1/\alpha^2)}$. Their techniques also yield an $2^{\text{poly}(1/\epsilon)}$-time test that determines whether the adjacency matrix of a graph can be changed in at most $\epsilon$ fraction of the locations to get a graph with high conductance. However, for the purpose of testing whether an $n$-vertex, $m$-edge graph is rapid mixing, we would need to approximate its conductance to within $\alpha = O(m/n^2)$; thus, only when $m = \Theta(n^2)$, would the algorithm in [27] run in $O(n)$ time.

We now discuss some other known results for testing of rapid mixing through eigenvalue computations. It is known that mixing [53, 38] is related to the separation between the two largest eigenvalues [4]. Standard techniques for approximating the eigenvalues of a dense $n \times n$ matrix run in $\Theta(n^3)$ floating-point operations and consume $\Theta(n^2)$ words of memory [33]. However, for a sparse $n \times n$ *symmetric* matrix with $m$ nonzero entries, $n \leq m$, "Lanczos algorithms" [46] accomplish the same task in $\Theta(n(m + \log n))$ floating-point operations, consuming $\Theta(n + m)$ storage. Furthermore, it is found in practice that these algorithms can be run for far fewer, even a constant number, of iterations while still obtaining highly accurate values for the outer and inner few eigenvalues.

**Streaming**  There is much work on the problem estimating the distance between distributions in data streaming models where space rather than time is limited (cf., [28, 3, 24, 26]). Another line of work [15] estimates the distance in frequency count distributions on words between various documents, where again space is limited. Guha et al. [35] have extended our result to estimating the closeness of distribution with respect to a range of $f$-divergences, which include $\ell_1$ distance. Testing distributions in streaming data models has been an active area of research in the recent years (see, for example, [11, 16, 36, 34, 17, 18, 13, 14]).

**Other Related Models** In an interactive setting, Sahai and Vadhan [52] show that, given distributions **p** and **q** generated by polynomial-size circuits, the problem of distinguishing whether **p** and **q** are close or far in $\ell_1$ norm is complete for statistical zero knowledge. Kannan and Yao [39] outlines a program checking framework for certifying the randomness of a program's output. In their model, one does not assume that samples from the input distribution are independent.

**Computational Learning Theory** There is a vast literature on testing statistical hypotheses. In these works, one is given examples chosen from the same distribution out of two possible choices, say **p** and **q**. The goal is to decide which of two distributions the examples are coming from. More generally, the goal can be stated as deciding which of two known classes of distributions contains the distribution generating the examples. This can be seen to be a generalization of our model as follows: Let the first class of distributions be the set of distributions of the form $\mathbf{q} \times \mathbf{q}$. Let the second class of distributions be the set of distributions of the form $\mathbf{q}_1 \times \mathbf{q}_2$ where the $\ell_1$ difference of $\mathbf{q}_1$ and $\mathbf{q}_2$ is at least $\epsilon$. Then, given examples from two distributions $\mathbf{p}_1, \mathbf{p}_2$, create a set of example pairs $(x, y)$ where $x$ is chosen according to $\mathbf{p}_1$ and $y$ according to $\mathbf{p}_2$ independently. Bounds and an optimal algorithm for the general problem for various distance measures are given in [19, 44, 20, 21, 41]. None of these give sublinear bounds in the domain size for our problem. The specific model of singleton hypothesis classes is studied by Yamanishi [56].

## 1.2 Notation

We use the following notation. We denote the set $\{1, \ldots, n\}$ with $[n]$. The notation $x \in_R [n]$ denotes that $x$ is chosen uniformly at random from the set $[n]$. The $\ell_1$ norm of a vector **v** is denoted by $\|\mathbf{v}\|_1$ and is equal to $\sum_{i=1}^{n} |v_i|$. Similarly, the $\ell_2$ norm is denoted by $\|\mathbf{v}\|_2$ and is equal to $\sqrt{\sum_{i=1}^{n} v_i^2}$, and $\|\mathbf{v}\|_\infty = \max_i |v_i|$. We assume our distributions are discrete distributions over $n$ elements, with labels in $[n]$, and will represent such a distribution as a vector $\mathbf{p} = (p_1, \ldots, p_n)$, where $p_i$ is the probability of outputting element $i$.

The *collision probability* of two distributions **p** and **q** is the probability that a sample from each of **p** and **q** yields the same element. Note that, for two distributions **p**, **q**, the collision probability is $\mathbf{p} \cdot \mathbf{q} = \sum_i p_i q_i$. To avoid ambiguity, we refer to the collision probability of **p** and **p** as the *self-collision probability* of **p**. Note that the self-collision probability of **p** is $\|\mathbf{p}\|_2^2$.

## 2 Testing Closeness of Distributions

The main goal of this section is to show how to test whether two distributions **p** and **q** are close in $\ell_1$ norm in sublinear time in the size of the domain of the distributions. We are given access to these distributions via black boxes which upon a query respond with an element of $[n]$ generated according to the respective distribution. Our main theorem is:

**Theorem 1** *Given parameters $\delta$ and $\epsilon$, and distributions $\mathbf{p}, \mathbf{q}$ over a set of $n$ elements, there is a test which runs in time $O(n^{2/3} \epsilon^{-8/3} \log(n/\delta))$ such that, if $\|\mathbf{p} - \mathbf{q}\|_1 \leq \max(\frac{\epsilon^{4/3}}{32 \sqrt[3]{n}}, \frac{\epsilon}{4\sqrt{n}})$, then the* test accepts with probability at least $1 - \delta$ and, if $\|\mathbf{p} - \mathbf{q}\|_1 > \epsilon$, then the test rejects with probability at least $1 - \delta$.

In order to prove this theorem, we give a test which determines whether $\mathbf{p}$ and $\mathbf{q}$ are close in $\ell_2$ norm. The test is based on estimating the self-collision and collision probabilities of $\mathbf{p}$ and $\mathbf{q}$. In particular, if $\mathbf{p}$ and $\mathbf{q}$ are close, one would expect that the self-collision probabilities of each are close to the collision probability of the pair. Formalizing this intuition, in Section 2.1, we prove:

**Theorem 2** *Given parameter $\delta$ and $\epsilon$, and distributions $\mathbf{p}$ and $\mathbf{q}$ over a set of $n$ elements, there exists a test such that, if $\|\mathbf{p} - \mathbf{q}\|_2 \leq \epsilon/2$, then the test accepts with probability at least $1 - \delta$ and, if $\|\mathbf{p} - \mathbf{q}\|_2 > \epsilon$, then the test rejects with probability at least $1 - \delta$. The running time of the test is $O(\epsilon^{-4} \log(1/\delta))$.*

The test used to prove Theorem 2 is given below in Figure 1. The number of pairwise self-collisions in multiset $F \subseteq [n]$ is the count of $i < j$ such that the $i$th sample in $F$ is same as the $j$th sample in $F$. Similarly, the number of collisions between $Q_p \subseteq [n]$ and $Q_q \subseteq [n]$ is the count of $(i, j)$ such that the $i$th sample in $Q_p$ is same as the $j$th sample in $Q_q$.

$\ell_2$-**Distance-Test**$(\mathbf{p}, \mathbf{q}, m, \epsilon, \delta)$
Repeat $O(\log(1/\delta))$ times

1. Let $F_p$ and $F_q$ be multisets of $m$ samples from $\mathbf{p}$ and $\mathbf{q}$, respectively. Let $r_p$ and $r_q$ be the numbers of pairwise self-collisions in $F_p$ and $F_q$, respectively.

2. Let $Q_p$ and $Q_q$ be multisets of $m$ samples from $\mathbf{p}$ and $\mathbf{q}$, respectively. Let $s_{pq}$ be the number of collisions between $Q_p$ and $Q_q$.

3. Let $r = \frac{2m}{m-1}(r_p + r_q)$. Let $s = 2s_{pq}$.

4. If $r - s > 3m^2\epsilon^2/4$, then reject the current iteration.

Reject if the majority of iterations reject, accept otherwise.

Figure 1: Algorithm $\ell_2$-Distance-Test

We use the parameter $m$ to indicate the number of samples needed by the test to get constant confidence. In order to bound the $\ell_2$ distance between $\mathbf{p}$ and $\mathbf{q}$ by $\epsilon$, setting $m = O(\frac{1}{\epsilon^4})$ suffices. By maintaining arrays which count the numbers of times, for example, $N_p(i)$ for $F_p$, that each element $i$ is sampled and summing $\binom{N_p(i)}{2}$ over all sampled $i$ in the domain, one can achieve the claimed running time bounds for computing an estimate of the collision probability. In this way, essentially $m^2$ estimations of the collision probability can be performed in $O(m)$ time.

Since $\|v\|_1 \leq \sqrt{n} \cdot \|v\|_2$, a simple way to extend the above test to an $L_1$ distance test is by setting $\epsilon' = \epsilon/\sqrt{n}$. This would give the correct output behavior for the tester. Unfortunately, due to the order of the dependence on $\epsilon$ in the $\ell_2$ distance test, the resulting running time is quadratic in $n$. It is possible, though, to achieve sublinear running times if the input distributions are known to be reasonably evenly distributed. We make this precise by a closer analysis of the variance of the estimator in the test in Lemma 5. In particular, we analyze the dependence of the variances of $s$ and $r$ on the parameter $b = \max(\|\mathbf{p}\|_\infty, \|\mathbf{q}\|_\infty)$. There we show that given $\mathbf{p}$ and $\mathbf{q}$ such that $b = O(n^{-\alpha})$, one can call $\ell_2$-**Distance-Test** with an error parameter of $\frac{\epsilon}{\sqrt{n}}$ and achieve running time of $O(\epsilon^{-4}(n^{1-\alpha/2} + n^{2-2\alpha}))$. Thus, when the maximum probability of any element is bounded, the $\ell_2$ distance test can in fact yield a sublinear-time algorithm for testing closeness in $L_1$ distance.

In the previous paragraph, we have noted that, for distributions with a bound on the maximum probability of any element, it is possible to test closeness with time and queries sublinear in the domain size. On the other hand, when the *minimum* probability element is quite large, the naive approach that we referred to in the introduction can be significantly more efficient. This suggests a *filtering* algorithm, which separates the domain of the distributions being tested into two parts – the *big* elements, or those elements to which the distributions assign relatively high probability weight, and the *small* elements, which are all other elements. Then, the naive tester is applied to the distributions restricted to the big elements, and the tester that is based on estimating the $\ell_2$ distance is applied to the distributions restricted to the small elements.

More specifically, we use the following definition to identify the elements with large weights.

**Definition 3 (Big element)** *An element $i$ is called* big with respect to a distribution $\mathbf{p}$ *if $p_i > (\epsilon/n)^{2/3}$*.

The complete test is given below in Figure 2. The proof of Theorem 1 is presented in Section 2.2.

$\ell_1$-**Distance-Test**$(\mathbf{p}, \mathbf{q}, \epsilon, \delta)$

1. Let $b = (\epsilon/n)^{2/3}$.
2. Sample $\mathbf{p}$ and $\mathbf{q}$ for $M = O(\epsilon^{-8/3} n^{2/3} \log(n/\delta))$ times.
3. Let $S^{\mathbf{p}}$ and $S^{\mathbf{q}}$ be the sample sets obtained from $\mathbf{p}$ and $\mathbf{q}$, respectively, by discarding elements that occur less than $(1 - \epsilon/26)Mb$ times.
4. If $S^{\mathbf{p}}$ and $S^{\mathbf{q}}$ are empty,
   $\ell_2$-**Distance-Test**$(\mathbf{p}, \mathbf{q}, O(n^{2/3}/\epsilon^{8/3}), \frac{\epsilon}{2\sqrt{n}}, \delta/2)$
   else
     i. Let $\ell_i^{\mathbf{p}}$ (resp., $\ell_i^{\mathbf{q}}$) be the times element $i$ appears in $S^{\mathbf{p}}$ (resp., $S^{\mathbf{q}}$).
     ii. Reject if $\sum_{i \in S^{\mathbf{p}} \cup S^{\mathbf{q}}} |\ell_i^{\mathbf{p}} - \ell_i^{\mathbf{q}}| > \epsilon M/8$.
     iii. Define $\mathbf{p}'$ as follows: Sample an element from $\mathbf{p}$. If this sample is not in $S^{\mathbf{p}} \cup S^{\mathbf{q}}$, output it; otherwise, output an $x \in_R [n]$. Define $\mathbf{q}'$ similarly.
     iv. $\ell_2$-**Distance-Test**$(\mathbf{p}', \mathbf{q}', O(n^{2/3}/\epsilon^{8/3}), \frac{\epsilon}{2\sqrt{n}}, \delta/2)$

Figure 2: Algorithm $\ell_1$-Distance-Test

## 2.1 Closeness in $\ell_2$ Norm

In this section, we analyze Algorithm $\ell_2$-**Distance-Test** and prove Theorem 2. The statistics $r_p$, $r_q$ and $s$ in Algorithm $\ell_2$-**Distance-Test** are estimators for the self-collision probability of $\mathbf{p}$, of $\mathbf{q}$, and of the collision probability between $\mathbf{p}$ and $\mathbf{q}$, respectively. If $\mathbf{p}$ and $\mathbf{q}$ are statistically close, we expect that the self-collision probabilities of each are close to the collision probability of the pair. These probabilities are exactly the inner products of these vectors. In particular, if the set $F_p$ of samples from $\mathbf{p}$ is given by $\{F_p^1, \ldots, F_p^m\}$, then, for any pair $i, j \in [m], i \neq j$, we have that $\Pr\left[F_p^i = F_p^j\right] = \mathbf{p} \cdot \mathbf{p} = \|\mathbf{p}\|_2^2$. By combining these statistics, we show that $r - s$ is an estimator for the desired value $\|\mathbf{p} - \mathbf{q}\|_2^2$.

In order to analyze the number of samples required to estimate $r - s$ to a high enough accuracy, we must also bound the variance of the variables $s$ and $r$ used in the test. One distinction to make between self-collisions and collisions between $\mathbf{p}$ and $\mathbf{q}$ is that, for the self-collisions, we only consider samples for which $i \neq j$, but this is not necessary for the collisions between $\mathbf{p}$ and $\mathbf{q}$. We accommodate this in our algorithm by scaling $r_p$ and $r_q$ appropriately. By this scaling and from the above discussion we see that $\mathrm{E}\,[s] = 2m^2(\mathbf{p} \cdot \mathbf{q})$ and that $\mathrm{E}\,[r - s] = m^2(\|\mathbf{p}\|_2^2 + \|\mathbf{q}\|_2^2 - 2(\mathbf{p} \cdot \mathbf{q})) = m^2(\|\mathbf{p} - \mathbf{q}\|_2^2)$.

A complication which arises from this scheme is that the pairwise samples are not independent. We use Chebyshev's inequality (see Appendix A) to bound the quality of the approximation, which in turn requires that we give a bound on the variance, as we do in this section.

Our techniques extend the work of Goldreich and Ron [32], where self-collision probabilities are used to estimate $\ell_2$ norm of a vector, and in turn the deviation of a distribution from uniform. In particular, their work provides an analysis of the statistics $r_p$ and $r_q$ above through the following lemma.

**Lemma 4 ([32])** *Consider the random variable $r_p$ in Algorithm $\ell_2$-**Distance-Test**. Then, $\mathrm{E}\,[r_p] = \binom{m}{2} \cdot \|\mathbf{p}\|_2^2$ and $\mathrm{Var}\,(r_p) \leq 2(\mathrm{E}\,[A])^{3/2}$.*

We next present a tighter variance bound given in terms of the largest weight in $\mathbf{p}$ and $\mathbf{q}$.

**Lemma 5** *There is a constant $c$ such that*

$$\mathrm{Var}\,(r_p) \leq m^2\|\mathbf{p}\|_2^2 + m^3\|\mathbf{p}\|_3^3 \leq c(m^3 b^2 + m^2 b),$$
$$\mathrm{Var}\,(r_q) \leq m^2\|\mathbf{q}\|_2^2 + m^3\|\mathbf{q}\|_3^3 \leq c(m^3 b^2 + m^2 b), \text{ and}$$
$$\mathrm{Var}\,(s) \leq c(m^3 b^2 + m^2 b),$$

*where $b = \max(\|\mathbf{p}\|_\infty, \|\mathbf{q}\|_\infty)$.*

PROOF: Let $F$ be the set $\{1, \ldots, m\}$. For $(i,j) \in F \times F$, define the indicator variable $C_{i,j} = 1$ if the $i$th element of $Q_p$ and the $j$th element of $Q_q$ are the same. Then, the variable from the algorithm $s_{pq} = \sum_{i,j} C_{i,j}$. Also define the notation $\bar{C}_{i,j} = C_{i,j} - \mathrm{E}\,[C_{i,j}]$. Given these definitions, we can write

$$\begin{aligned}
\mathrm{Var}\left(\sum_{(i,j) \in F \times F} C_{i,j}\right) &= \mathrm{E}\left[\left(\sum_{(i,j) \in F \times F} \bar{C}_{i,j}\right)^2\right] \\
&= \mathrm{E}\left[\sum_{(i,j) \in F \times F} (\bar{C}_{i,j})^2 + 2 \sum_{(i,j) \neq (k,l) \in F \times F} \bar{C}_{i,j}\bar{C}_{k,l}\right] \\
&\leq \mathrm{E}\left[\sum_{(i,j) \in F \times F} C_{i,j}\right] + 2 \cdot \mathrm{E}\left[\sum_{(i,j) \neq (k,l) \in F \times F} \bar{C}_{i,j}\bar{C}_{k,l}\right] \\
&= m^2(\mathbf{p} \cdot \mathbf{q}) + 2 \cdot \mathrm{E}\left[\sum_{(i,j) \neq (k,l) \in F \times F} \bar{C}_{i,j}\bar{C}_{k,l}\right]
\end{aligned}$$

To analyze the last expectation, we use two facts. First, it is easy to see, by the definition of covariance, that $\mathrm{E}\,[\bar{C}_{i,j}\bar{C}_{k,l}] \leq \mathrm{E}\,[C_{i,j}C_{k,l}]$. Secondly, we note that $C_{i,j}$ and $C_{k,l}$ are not independent only when $i = k$ or $j = l$. Expanding the sum, we get

$$
\begin{aligned}
\mathrm{E}\left[\sum_{\substack{(i,j),(k,l) \in F \times F \\ (i,j) \neq (k,l)}} \bar{C}_{i,j} \bar{C}_{k,l}\right] &= \mathrm{E}\left[\sum_{\substack{(i,j),(i,l) \in F \times F \\ j \neq l}} \bar{C}_{i,j} \bar{C}_{i,l} + \sum_{\substack{(i,j),(k,j) \in F \times F \\ i \neq k}} \bar{C}_{i,j} \bar{C}_{k,j}\right] \\
&\leq \mathrm{E}\left[\sum_{\substack{(i,j),(i,l) \in F \times F \\ j \neq l}} C_{i,j} C_{i,l} + \sum_{\substack{(i,j),(k,j) \in F \times F \\ i \neq k}} C_{i,j} C_{k,j}\right] \\
&\leq cm^3 \sum_{\ell \in [n]} p_\ell q_\ell^2 + p_\ell^2 q_\ell \leq cm^3 b^2 \sum_{\ell \in [n]} q_\ell \leq cm^3 b^2
\end{aligned}
$$

for some constant $c$. Next, we bound $\mathrm{Var}\,(r)$ similarly to $\mathrm{Var}\,(s)$ using the argument in the proof of Lemma 4 from [32]. Consider an analogous calculation to the preceding inequality for $\mathrm{Var}\,(r_p)$ (similarly, for $\mathrm{Var}\,(r_q)$) where $X_{ij} = 1$ for $1 \leq i < j \leq m$ if the $i$th and $j$th samples in $F_p$ are the same. Similarly to above, define $\bar{X}_{ij} = X_{ij} - \mathrm{E}\,[X_{ij}]$. Then, we get

$$
\begin{aligned}
\mathrm{Var}\,(r_p) &= \mathrm{E}\left[\left(\sum_{1 \leq i < j \leq m} \bar{X}_{ij}\right)^2\right] \\
&= \sum_{1 \leq i < j \leq m} \mathrm{E}\left[\bar{X}_{i,j}^2\right] + 4 \sum_{1 \leq i < j < k \leq m} \mathrm{E}\left[\bar{X}_{i,j} \bar{X}_{i,k}\right] \\
&\leq \binom{m}{2} \cdot \sum_{t \in [n]} p_t^2 + 4 \cdot \binom{m}{3} \sum_{t \in [n]} p_t^3 \\
&\leq O(m^2) \cdot b + O(m^3) \cdot b^2.
\end{aligned}
$$

Thus, we get the upper bound for both variances. □

**Corollary 6** *There is a constant $c$ such that $\mathrm{Var}\,(r - s) \leq c(m^3 b^2 + m^2 b)$, where $b = \max(\|\mathbf{p}\|_\infty, \|\mathbf{q}\|_\infty)$.*

PROOF: Since variance is additive for independent random variables, we get $\mathrm{Var}\,(r - s) \leq c(m^3 b^2 + m^2 b)$. □

Now using Chebyshev's inequality, it follows that if we choose $m = O(\epsilon^{-4})$, we can achieve an error probability less than $1/3$. It follows from standard techniques that with $O(\log \frac{1}{\delta})$ iterations we can achieve an error probability at most $\delta$.

Finally, we can analyze the behavior of the algorithm.

**Theorem 7** *For two distributions $\mathbf{p}$ and $\mathbf{q}$ such that $b = \max(\|\mathbf{p}\|_\infty, \|\mathbf{q}\|_\infty)$ and $m = \Omega((b^2 + \epsilon^2 \sqrt{b})/\epsilon^4)$, if $\|\mathbf{p} - \mathbf{q}\|_2 \leq \epsilon/2$, then $\ell_2$-**Distance-Test**$(\mathbf{p}, \mathbf{q}, m, \epsilon, \delta)$ accepts with probability at least $1 - \delta$. If $\|\mathbf{p} - \mathbf{q}\|_2 > \epsilon$ then $\ell_2$-**Distance-Test**$(\mathbf{p}, \mathbf{q}, m, \epsilon, \delta)$ accepts with probability less than $\delta$. The running time is $O(m \log(1/\delta))$.*

PROOF: For our statistic $A = (r - s)$, we can say, using Chebyshev's inequality and Corollary 6, that for some constant $c$,

$$\Pr\left[|A - \mathrm{E}\,[A]| > \rho\right] \leq \frac{c(m^3 b^2 + m^2 b)}{\rho^2}.$$

Recalling that $\mathrm{E}[A] = m^2(\|\mathbf{p} - \mathbf{q}\|_2^2)$, we observe that the $\ell_2$-**Distance-Test** can distinguish between the cases $\|\mathbf{p} - \mathbf{q}\|_2 \leq \epsilon/2$ and $\|\mathbf{p} - \mathbf{q}\|_2 > \epsilon$ if $A$ is within $m^2\epsilon^2/4$ of its expectation. We can bound the error probability by

$$\Pr\left[|A - \mathrm{E}[A]| > m^2\epsilon^2/4\right] \leq \frac{16c(m^3b^2 + m^2b)}{m^4\epsilon^4}.$$

Thus, for $m = \Omega((b^2 + \epsilon^2\sqrt{b})/\epsilon^4)$, the probability above is bounded by a constant. This error probability can be reduced to $\delta$ by $O(\log(1/\delta))$ repetitions. □

## 2.2 Closeness in $L_1$ Norm

The $\ell_1$-**Distance-Test** proceeds in two phases. The first phase of the algorithm (lines 1–3 and 4(i)–(ii)) determines which elements of the domain are the big elements (as defined in Definition 3) and estimates their contribution to the distance $\|\mathbf{p} - \mathbf{q}\|_1$. The second phase (lines 4(iii)–(iv)) filters out the big elements and invokes the $\ell_2$-**Distance-Test** on the filtered distribution with closeness parameter $\epsilon/(2\sqrt{n})$. The correctness of this subroutine call is given by Theorem 7 with $b = 2\epsilon^{2/3}n^{-2/3}$. With these substitutions, the number of samples $m$ is $O(\epsilon^{-8/3}n^{2/3})$. The choice of threshold $b$ in $\ell_1$-**Distance-Test** for the weight of the big elements arises from optimizing the running-time trade-off between the two phases of the algorithm.

We need to show that by using a sample of size $O(\epsilon^{-8/3}n^{2/3}\log(n/\delta))$, we can estimate the weights of each of the big elements to within a multiplicative factor of $1 + O(\epsilon)$, with probability at least $1 - \delta/2$.

**Lemma 8** *Let $b = \epsilon^{2/3}n^{-2/3}$. In $\ell_1$-Distance-Test, after taking $M = O(\frac{n^{2/3}\log(n/\delta)}{\epsilon^{8/3}})$ samples from a distribution $\mathbf{p}$, we define $\bar{p}_i = \ell_i^\mathbf{p}/M$. Then, with probability at least $1 - \delta/2$, the following hold for all $i$: (1) if $p_i \geq (1 - \epsilon/13)b$, then $|\bar{p}_i - p_i| < \frac{\epsilon}{26}\max(p_i, b)$, (2) if $p_i < (1 - \epsilon/13)b$, then $\bar{p}_i < (1 - \epsilon/26)b$.*

PROOF: We analyze two cases; we use Chernoff bounds to show that, for each $i$, the following holds: If $p_i > b$, then

$$\Pr[|\bar{p}_i - p_i| > \epsilon p_i/26] < \exp(-O(\epsilon^2 M p_i)) < \exp(-O(\epsilon^2 Mb)) \leq \frac{\delta}{2n}.$$

If $p_i \leq b$, then

$$\Pr[|\bar{p}_i - p_i| > \epsilon b/26] \leq \Pr\left[|\bar{p}_i - p_i| > \frac{\epsilon b}{26 p_i}p_i\right] < \exp(-O(\epsilon^2 b^2 M/p_i)) \leq \exp(-O(\epsilon^2 Mb)) \leq \frac{\delta}{2n}.$$

The lemma follows by the union bound. □

Now we are ready to prove our main theorem.

**Theorem 9** *For $\epsilon \geq 1/\sqrt{n}$, $\ell_1$-Distance-Test accepts distributions $\mathbf{p}, \mathbf{q}$ such that $\|\mathbf{p} - \mathbf{q}\|_1 \leq \max(\frac{\epsilon^{4/3}}{32\sqrt[3]{n}}, \frac{\epsilon}{4\sqrt{n}})$, and rejects when $\|\mathbf{p} - \mathbf{q}\|_1 > \epsilon$, with probability at least $1 - \delta$. The running time of the test is $O(\epsilon^{-8/3}n^{2/3}\log(n/\delta))$.*

PROOF: Suppose items (1) and (2) from Lemma 8 hold for all $i$, and for both $\mathbf{p}$ and $\mathbf{q}$. By Lemma 8, this event happens with probability at least $1 - \delta/2$.

Let $S = S^{\mathbf{p}} \cup S^{\mathbf{q}}$. By our assumption, all the big elements of both $\mathbf{p}$ and $\mathbf{q}$ are in $S$, and no element that has weight less than $(1-\epsilon/13)b$ in both distributions is in $S$. Let $\Delta_1$ be the $\ell_1$ distance attributed to the elements in $S$; that is, $\sum_{i \in S} |p_i - q_i|$. Let $\Delta_2 = \|\mathbf{p}' - \mathbf{q}'\|_1$ (in the case that $S$ is empty, $\Delta_1 = 0$, $\mathbf{p} = \mathbf{p}'$ and $\mathbf{q} = \mathbf{q}'$). Note that $\Delta_1 \leq \|\mathbf{p} - \mathbf{q}\|_1$. We can show that $\Delta_2 \leq \|\mathbf{p} - \mathbf{q}\|_1$, and $\|\mathbf{p} - \mathbf{q}\|_1 \leq 2\Delta_1 + \Delta_2$.

Next, we show that the algorithm estimates $\Delta_1$ in a brute-force manner to within an additive error of $\epsilon/9$. By Lemma 8, the error on the $i$th term of the sum is bounded by

$$\frac{\epsilon}{26}(\max(p_i, b) + \max(q_i, b)) \leq \frac{\epsilon}{26}(p_i + q_i + 2\epsilon b/13),$$

where the last inequality follows from that $p_i$ and $q_i$ are at least $(1 - \epsilon/13)b$. Consider the sum over $i$ of these error terms. Notice that this sum is over at most $2/((1 - \epsilon/13)b)$ elements in $S$. Hence, the total additive error is bounded by

$$\sum_{i \in S} \frac{\epsilon}{26}(p_i + q_i + 2\epsilon b/13) \leq \frac{\epsilon}{26}(2 + 4\epsilon/(13 - \epsilon)) \leq \epsilon/9$$

since $\epsilon \leq 2$.

Note that $\max(\|\mathbf{p}'\|_\infty, \|\mathbf{q}'\|_\infty) \leq b + n^{-1} \leq 2b$ for $\epsilon \geq 1/\sqrt{n}$. So, we can use the $\ell_2$-**Distance-Test** on $\mathbf{p}'$ and $\mathbf{q}'$ with $m = O(\epsilon^{-8/3} n^{2/3})$ as shown by Theorem 7.

If $\|\mathbf{p} - \mathbf{q}\|_1 < \frac{\epsilon^{4/3}}{32 \sqrt[3]{n}}$, then so are $\Delta_1$ and $\Delta_2$. The first phase of the algorithm clearly accepts. Using the fact that, for any vector $v$, $\|v\|_2^2 \leq \|v\|_1 \cdot \|v\|_\infty$, we get $\|\mathbf{p}' - \mathbf{q}'\|_2 \leq \frac{\epsilon}{4\sqrt{n}}$. Therefore, the $\ell_2$-**Distance-Test** accepts with probability at least $1 - \delta/2$. Similarly, if $\|\mathbf{p} - \mathbf{q}\|_1 > \epsilon$, then either $\Delta_1 > \epsilon/4$ or $\Delta_2 > \epsilon/2$. Either the first phase of the algorithm or the $\ell_2$-**Distance-Test** will reject.

To see the running time bound, note that the time for the first phase is $O(n^{2/3} \epsilon^{-8/3} \log(n/\delta))$ and that the time for $\ell_2$-**Distance-Test** is $O(n^{2/3} \epsilon^{-8/3} \log \frac{1}{\delta})$. It is easy to see that our algorithm makes an error either when it makes a bad estimation of $\Delta_1$ or when $\ell_2$-**Distance-Test** makes an error. So, the probability of error is bounded by $\delta$. □

The next theorem improves this result by looking at the dependence of the variance calculation in Section 2.1 on $L_\infty$ norms of the distributions separately.

**Theorem 10** *Given two black-box distributions $\mathbf{p}, \mathbf{q}$ over $[n]$, with $\|\mathbf{p}\|_\infty \leq \|\mathbf{q}\|_\infty$, there is a test requiring $O((n^2 \|\mathbf{p}\|_\infty \|\mathbf{q}\|_\infty \epsilon^{-4} + n\sqrt{\|\mathbf{q}\|_\infty} \epsilon^{-2}) \log(1/\delta))$ samples that (1) if $\|\mathbf{p} - \mathbf{q}\|_1 \leq \frac{\epsilon^2}{\sqrt[3]{n}}$, it accepts with probability at least $1 - \delta$ and (2) if $\|\mathbf{p} - \mathbf{q}\|_1 > \epsilon$, it rejects with probability at least $1 - \delta$.*

## 2.3 Testing $\ell_1$ Distance from Uniformity

A special case of Theorem 2 gives a constant-time algorithm which provides an additive approximation of the $\ell_2$ distance of a distribution from the uniform distribution. For the problem of testing that $\mathbf{p}$ is close to the uniform distribution in $\ell_1$ distance (i.e., testing closeness when $\mathbf{q}$ is the uniform distribution), one can get a better sample complexity dependence on $n$.

**Theorem 11** *Given $\epsilon \leq 1$ and a black-box distribution $\mathbf{p}$ over $[n]$, there is a test that takes $O(\epsilon^{-4} \cdot \sqrt{n} \cdot \log(1/\delta))$ samples, accepts with probability at least $1 - \delta$ if $\|\mathbf{p} - U_{[n]}\|_1 \leq \epsilon/\sqrt{3n}$, and rejects with probability at least $1 - \delta$ if $\|\mathbf{p} - U_{[n]}\|_1 > \epsilon$.*

The proof of Theorem 11 relies on the following lemma, which can be proven using techniques from Goldreich and Ron [32] (see also Lemma 5 in this paper).

**Lemma 12** *Given a black-box distribution $\mathbf{p}$ over $[n]$, there is an algorithm that takes $O(\epsilon^{-2} \cdot \sqrt{n} \cdot \log(1/\delta))$ samples and estimates $\|\mathbf{p}\|_2^2$ within an error of $\epsilon\|\mathbf{p}\|_2^2$, with probability at least $1 - \delta$.*

PROOF: [of Lemma 12] Consider the random variable $r_p$ from the $\ell_2$-Distance-Test. Since $\mathrm{E}[r_p] = \binom{m}{2} \cdot \|\mathbf{p}\|_2^2$, we only need to show that it does not deviate from its expectation too much with high probability. Again, using Chebyshev's inequality and Lemma 5,

$$\Pr\left[|r_p - \mathrm{E}[r_p]| > \epsilon \mathrm{E}[r_p]\right] \leq \frac{O(m^2 \|\mathbf{p}\|_2^2 + m^3 \|\mathbf{p}\|_2^3)}{\epsilon^2 m^4 \|\mathbf{p}\|_2^4} \leq \frac{1}{4},$$

where the last inequality follows for $m = O(\epsilon^{-2}\sqrt{n})$ from the fact that $\|\mathbf{p}\|_2 \geq n^{-1/2}$. The confidence can be boosted to $1 - \delta$ using $O(\log(1/\delta))$ repetitions. □

We note that, for an additive approximation of $\|\mathbf{p}\|_2$, an analogous argument to the proof above will yield an algorithm that uses $O(\epsilon^{-4})$ samples.

PROOF: [of Theorem 11] The algorithm, given in Figure 3, estimates $\|\mathbf{p}\|_2^2$ within $\epsilon^2 \|\mathbf{p}\|_2^2/5$ using the algorithm from Lemma 12 and accepts only if the estimate is below $(1 + 3\epsilon^2/5)/n$.

**Uniformity-Distance-Test$(\mathbf{p}, m, \epsilon, \delta)$**

1. Accept if GR-Uniformity-$\ell_2$-Distance-Test$(\mathbf{p}, \epsilon^2/5)$ returns an estimate at most $(1 + 3\epsilon^2/5)/n$.

2. Otherwise, reject.

Figure 3: Algorithm Uniformity-Distance-Test

First, observe the following relationship between the $\ell_2$ distance to the uniform distribution and the collision probability.

$$\|\mathbf{p} - U_{[n]}\|_2^2 = \sum_i (p_i - \frac{1}{n})^2 = \sum p_i^2 - \frac{2}{n} \cdot \sum p_i + \frac{1}{n} = \|\mathbf{p}\|_2^2 - \frac{1}{n} \quad (1)$$

If $\|\mathbf{p} - U_{[n]}\|_1 \leq \epsilon/\sqrt{3n}$, then $\|\mathbf{p} - U_{[n]}\|_2^2 \leq \epsilon^2/3n$. Using (1), we see that $\|\mathbf{p}\|_2^2 \leq (1 + \epsilon^2/3)/n$. Hence, for $\epsilon \leq 1$, the estimate will be below $(1 + \epsilon^2/5)(1 + \epsilon^2/3)/n \leq (1 + 3\epsilon^2/5)/n$ with probability at least $1 - \delta$.

Conversely, suppose the estimate of $\|\mathbf{p}\|_2^2$ is below $(1 + 3\epsilon^2/5)/n$. By Lemma 12, $\|\mathbf{p}\|_2^2 \leq (1 + 3\epsilon^2/5)/((1 - \epsilon^2/5)n) \leq (1 + \epsilon^2)/n$ for $\epsilon \leq 1$. Therefore, by (1), we can write

$$\|\mathbf{p} - U_{[n]}\|_2^2 = \|\mathbf{p}\|_2^2 - \frac{1}{n} \leq \epsilon^2/n.$$

So, we have $\|\mathbf{p} - U_{[n]}\|_2 \leq \epsilon/\sqrt{n}$. Finally, by the relation between $\ell_1$ and $\ell_2$ norms, $\|\mathbf{p} - U_{[n]}\|_1 \leq \epsilon$.

The sample complexity of the procedure will be $O(\epsilon^{-4} \cdot \sqrt{n} \cdot \log(1/\delta))$, arising from the estimation of $\|\mathbf{p}\|_2^2$ within $\epsilon^2 \|\mathbf{p}\|_2^2/5$. □

# 3 Lower Bounding the Sample Complexity

In this section we consider lower bounds on the sample complexity of testing closeness of distributions. In a previous version of this paper [9], we claimed an almost matching $\Omega(n^{2/3})$ lower bound on the sample complexity for testing the closeness of two arbitrary distributions. Although it was later determined that there were gaps in the proofs, recent results of [54] have shown that the in fact the almost matching lower bounds do hold. Although new proof techniques were needed, certain technical ideas such as "Poissonization" and the characterization of "canonical forms of testing algorithms" that first appeared in the earlier version of this work did in fact turn out to be useful in the correct lower bound proof of [54]. We will outline those ideas in this section.

We begin by discussing a characterization of canonical algorithms for testing properties of distributions. Then we describe a pair of families of distributions that were suggested in the earlier version of this work, and were in fact used by Valiant [54] in showing the correct lower bound. Next, we investigate the required dependence on $\epsilon$. Finally, we briefly consider naive learning algorithms, which can be defined as algorithms that, given samples from a distribution, output a distribution with small distance to the input distribution. We show that naive learning algorithms require $\Omega(n)$ samples. We also note that, more recently, the dependency of testing uniformity on distance parameter $\epsilon$ and $n$ has been tightly characterized to be $\Theta(\sqrt{n}/\epsilon^2)$ by Paninski [45].

## 3.1 Characterization of Canonical Algorithms for Testing Properties of Distributions

In this section, we characterize canonical algorithms for testing properties of distributions defined by permutation-invariant functions. The argument hinges on the irrelevance of the labels of the domain elements for such a function. We obtain this canonical form in two steps, corresponding to the two lemmas below. The first step makes explicit the intuition that such an algorithm should be symmetric, that is, the algorithm would not benefit from discriminating among the labels. In the second step, we remove the use of labels altogether, and show that we can present the sample to the algorithm in an aggregate fashion. Raskhodnikova et al. [48] use this chararecterization of canonical algorithms for proving lower bounds on the sample complexity of distribution support size and element distinctness problems.

Characterizations of property testing algorithms have been studied in other settings. For example, using similar techniques, Alon et al. [2] show a canonical form for algorithms for testing graph properties. Later, Goldreich and Trevisan [29] formally prove the result by Alon et al. In a different setting, Bar-Yossef et al. [6] show a canonical form for sampling algorithms that approximate symmetric functions of the form $f : A^n \to B$ where $A$ and $B$ are arbitrary sets. In the latter setting, the algorithm is given oracle access to the input vector and takes samples from the coordinate values of this vector.

Next, we give the definitions of basic concepts on which we build a characterization of canonical algorithms for testing properties of distributions. Then, we describe and prove our characterization.

**Definition 13 (Permutation of a distribution)** *For a distribution $\mathbf{p}$ over $[n]$ and a permutation $\pi$ on $[n]$, define $\pi(\mathbf{p})$ to be the distribution such that for all $i$, $\pi(\mathbf{p})_{\pi(i)} = p_i$.*

**Definition 14 (Symmetric Algorithm)** *Let $\mathcal{A}$ be an algorithm that takes samples from $k$ discrete black-box distributions over $[n]$ as input. We say that $\mathcal{A}$ is* symmetric *if, once the distributions are fixed, the output distribution of $\mathcal{A}$ is identical for any permutation of the distributions.*

**Definition 15 (Permutation-invariant function)** *A $k$-ary function $f$ on distributions over $[n]$ is* permutation-invariant *if for any permutation $\pi$ on $[n]$, and all distributions $(\mathbf{p}^{(1)}, \ldots, \mathbf{p}^{(k)})$,*

$$f(\mathbf{p}^{(1)}, \ldots, \mathbf{p}^{(k)}) = f(\pi(\mathbf{p}^{(1)}), \ldots, \pi(\mathbf{p}^{(k)})).$$

**Lemma 16** *Let $\mathcal{A}$ be an arbitrary testing algorithm for a $k$-ary property $\mathcal{P}$ defined by a permutation-invariant function. Suppose $\mathcal{A}$ has sample complexity $s(n)$, where $n$ is the domain size of the distributions. Then, there exists a symmetric algorithm that tests the same property of distributions with sample complexity $s(n)$.*

PROOF: Given the algorithm $\mathcal{A}$, construct a symmetric algorithm $\mathcal{A}'$ as follows: Choose a random permutation of the domain elements. Upon taking $s(n)$ samples, apply this permutation to each sample. Pass this (renamed) sample set to $\mathcal{A}$ and output according to $\mathcal{A}$.

It is clear that the sample complexity of the algorithm does not change. We need to show that the new algorithm also maintains the testing features of $\mathcal{A}$. Suppose that the input distributions $(\mathbf{p}^{(1)}, \ldots, \mathbf{p}^{(k)})$ have the property $\mathcal{P}$. Since the property is defined by a permutation-invariant function, any permutation of the distributions maintains this property. Therefore, the permutation of the distributions should be accepted as well. Let $S_n$ denote the set of all permutations on $[n]$. Then,

$$\Pr\left[\mathcal{A}' \text{ accepts } (\mathbf{p}^{(1)}, \ldots, \mathbf{p}^{(k)})\right] = \sum_{\pi \in S_n} \frac{1}{n!} \Pr\left[\mathcal{A} \text{ accepts } (\pi(\mathbf{p}^{(1)}), \ldots, \pi(\mathbf{p}^{(k)}))\right],$$

which is at least $2/3$ by the accepting probability of $\mathcal{A}$.

An analogous argument on the failure probability for the case of the distributions $(\mathbf{p}^{(1)}, \ldots, \mathbf{p}^{(k)})$ that should be rejected completes the proof. $\square$

In order to avoid introducing additional randomness in $\mathcal{A}'$, we can try $\mathcal{A}$ on all possible permutations and output the majority vote. This change would not affect the sample complexity, and it can be shown that it maintains correctness.

**Definition 17 (Fingerprint of a sample)** *Let $S_1$ and $S_2$ be multisets of at most $s$ samples taken from two black-box distributions over $[n]$, $\mathbf{p}$ and $\mathbf{q}$, respectively. Let the random variable $C_{ij}$, for $0 \leq i, j \leq s$, denote the number of elements that appear exactly $i$ times in $S_1$ and exactly $j$ times in $S_2$. The collection of values that the random variables $\{C_{ij}\}_{0 \leq i,j \leq s}$ take is called the* fingerprint *of the sample.*

For example, let sample sets be $S_1 = \{5, 7, 3, 3, 4\}$ and $S_2 = \{2, 4, 3, 2, 6\}$. Then, $C_{10} = 2$ (elements 5 and 7), $C_{01} = 1$ (element 6), $C_{11} = 1$ (element 4), $C_{02} = 1$ (element 2), $C_{21} = 1$ (element 3), and for remaining $i, j$'s, $C_{ij} = 0$.

**Lemma 18** *If there exists a symmetric algorithm $\mathcal{A}$ for testing a binary property of distributions defined by a permutation-invariant function, then there exist an algorithm for the same task that gets as input only the fingerprint of the sample that $\mathcal{A}$ takes.*

PROOF: Fix a canonical order for $C_{ij}$'s in the fingerprint of a sample. Let us define the following transformation on the sample: Relabel the elements such that the elements that appear exactly the same number of times from each distribution (i.e., the ones that contribute to a single $C_{ij}$ in the fingerprint) have consecutive labels and the labels are grouped to conform to the canonical order of $C_{ij}$'s. Let us call this transformed sample the standard form of the sample. Since the algorithm $\mathcal{A}$ is symmetric and the property is defined by a permutation-invariant function, such a transformation does not affect the output of $\mathcal{A}$. So, we can further assume that we always present the sample to the algorithm in the standard form.

It is clear that given a sample, we can easily write down the fingerprint of the sample. Moreover, given the fingerprint of a sample, we can always construct a sample $(S_1, S_2)$ in the standard form using the following algorithm: (1) Initialize $S_1$ and $S_2$ to be empty, and $e = 1$, (2) for every $C_{ij}$ in the canonical order, and for $C_{ij} = k_{ij}$ times, include $i$ and $j$ copies of the element $e$ in $S_1$ and $S_2$, respectively, then increment $e$. This algorithm shows a one-to-one and onto correspondence between all possible sample sets in the standard form and all possible $\{C_{ij}\}_{0 \leq i,j \leq s}$ values.

Consider the algorithm $\mathcal{A}'$ that takes the fingerprint of a sample as input. Next, by using algorithm from above, algorithm $\mathcal{A}'$ constructs the sample in the standard form. Finally, $\mathcal{A}'$ outputs what $\mathcal{A}$ outputs on this sample. □

**Remark 19** *Note that the definition of the fingerprint from Definition 17 can be generalized for a collection of $k$ sample sets from $k$ distributions for any $k$. An analogous lemma to Lemma 18 can be proven for testing algorithms for $k$-ary properties of distributions defined by a permutation-invariant function. We fixed $k = 2$ for ease of notation.*

## 3.2 Towards a Lower Bound on the Sample Complexity of Testing Closeness

In this section, we present techniques that were later used by Valiant [54] to prove a lower bound on the sample complexity of testing closeness in $\ell_1$ distance as a function of the size $n$ of the domain of the distributions. We give a high-level description of the proof, indicate where our reasoning breaks down and where Valiant [54] comes in.

**Theorem 20 ([54])** *Given any algorithm using only $o(n^{2/3})$ samples from two discrete black-box distributions over $[n]$, for all sufficiently large $n$, there exist distributions $\mathbf{p}$ and $\mathbf{q}$ with $\ell_1$ distance 1 such that the algorithm will be unable to distinguish the case where one distribution is $\mathbf{p}$ and the other is $\mathbf{q}$ from the case where both distributions are $\mathbf{p}$.*

By Lemma 16, we may restrict our attention to symmetric algorithms. Fix a testing algorithm $\mathcal{A}$ that uses $o(n^{2/3})$ samples from each of the input distributions.

Let us assume, without loss of generality, that $n$ is a multiple of four and $n^{2/3}$ is an integer. We define the distributions $\mathbf{p}$ and $\mathbf{q}$ as follows: (1) For $1 \leq i \leq n^{2/3}$, $p_i = q_i = \frac{1}{2n^{2/3}}$. We call these elements the *heavy* elements. (2) For $n/2 < i \leq 3n/4$, $p_i = \frac{2}{n}$ and $q_i = 0$. We call these element the *light* elements of $\mathbf{p}$. (3) For $3n/4 < i \leq n$, $q_i = \frac{2}{n}$ and $p_i = 0$. We call these elements the *light* elements of $\mathbf{q}$. (4) For the remaining $i$'s, $p_i = q_i = 0$. Note that these distributions do not depend on $\mathcal{A}$.

The $\ell_1$ distance of $\mathbf{p}$ and $\mathbf{q}$ is 1. Now, consider the following two cases:

Case 1: The algorithm is given access to two black-box distributions: both of which output samples according to the distribution $\mathbf{p}$.

Case 2: The algorithm is given access to two black-box distributions: the first one outputs samples according to the distribution $\mathbf{p}$ and the second one outputs samples according to the distribution $\mathbf{q}$.

To get a sense of why these distributions should be hard for any distance testing algorithm, note that when restricted to the heavy elements, both distributions are identical. The only difference between **p** and **q** comes from the light elements, and the crux of the proof is to show that this difference will not change the relevant statistics in a statistically significant way. For example, consider the statistic which counts the number of elements that occur exactly once from each distribution. One would like to show that this statistic has a very similar distribution when generated by Case 1 and Case 2, because the expected number of such elements that are light is much less than the standard deviation of the number of such elements that are heavy.

Our initial attempts at formalizing the intuition above were incomplete. However, completely formalizing this intuition, Valiant [54] subsequently showed that a symmetric algorithm with sample complexity $o(n^{2/3})$ can not distinguish between these two cases. By Lemma 16, the theorem follows.

**Poissonization** For simplifying the proof, it would be useful to have the frequency of each element be independent of the frequencies of the other elements. To achieve this, we assume that algorithm $\mathcal{A}$ first chooses two integers $s_1$ and $s_2$ independently from a Poisson distribution with the parameter $\lambda = s = o(n^{2/3})$. The Poisson distribution with the positive parameter $\lambda$ has the probability mass function $p(k) = \exp(-\lambda)\lambda^k/k!$. Then, after taking $s_1$ samples from the first distribution and $s_2$ samples from the second distribution, $\mathcal{A}$ decides whether to accept or reject the distributions. In the following, we give an overview of the proof that $\mathcal{A}$ cannot distinguish between Case 1 and Case 2 with success probability at least $2/3$. Since both $s_1$ and $s_2$ will have values larger than $s/2$ with probability at least $1 - o(1)$ and the statistical distance of the distributions of two random variables (i.e., the distributions on the samples) is bounded, it will follow that no symmetric algorithm with sample complexity $s/2$ can.

Let $F_i$ be the random variable corresponding to the number of times the element $i$ appears in the sample from the first distribution. Define $G_i$ analogously for the second distribution. It is well known that $F_i$ is distributed identically to the Poisson distribution with parameter $\lambda = sr$, where $r$ is the probability of element $i$ (cf., Feller ([25], p. 216). Furthermore, it can also be shown that all $F_i$'s are mutually independent. Thus, the total number of samples from the heavy elements and the total number of samples from the light elements are independent.

**Canonical Testing Algorithms** Recall the definition of the fingerprint of a sample from Section 3.1. The random variable $C_{ij}$, denotes the number of elements that appear exactly $i$ times in the sample from the first distribution and exactly $j$ times in the sample from the second distribution. We can then assume that the algorithm is only given the fingerprint of the sample, and apply Lemma 18.

Arguing in this way can lead to several subtle pitfalls, which Valiant's proof [54] circumvents by developing a body of additional, very nontrivial, technical machinery to show that the distributions on the fingerprint when the samples come from Case 1 or Case 2 are indistinguishable.

### 3.3 Other Lower Bounds

In this section, we first give two lower bounds for the sample complexity of testing closeness in terms of the distance parameter $\epsilon$. Then, we show that a naive learning algorithm for distributions require $\Omega(n)$ samples.

By appropriately modifying the distributions **p** and **q** from the proof, we can give a stronger version of Theorem 20 with a dependence on $\epsilon$.

**Corollary 21** *Given any test using only $o(n^{2/3}/\epsilon^{2/3})$ samples, there exist distributions* **a** *and* **b** *of $\ell_1$ distance $\epsilon$ such that the test will be unable to distinguish the case where one distribution is* **a** *and the other is* **b** *from the case where both distributions are* **a**.

We can get a lower bound of $\Omega(\epsilon^{-2})$ for testing the $\ell_2$ distance with a rather simple proof.

**Theorem 22** *Given any test using only $o(\epsilon^{-2})$ samples, there exist distributions* **a** *and* **b** *of $\ell_2$ distance $\epsilon$ such that the test will be unable to distinguish the case where one distribution is* **a** *and the other is* **b** *from the case where both distributions are* **a**.

PROOF: Let $n = 2$, $a_1 = a_2 = 1/2$ and $b_1 = 1/2 - \epsilon/\sqrt{2}$ and $b_2 = 1/2 + \epsilon/\sqrt{2}$. Distinguishing these distributions is exactly the question of distinguishing a fair coin from a coin of bias $\Theta(\epsilon)$ which is well known to require $\Theta(\epsilon^{-2})$ coin flips. □

The next theorem shows that learning a distribution using sublinear number of samples is not possible.

**Theorem 23** *Suppose we have an algorithm that draws $o(n)$ samples from some unknown distribution* **b** *and outputs a distribution* **c**. *There is some distribution* **b** *for which the output* **c** *is such that* **b** *and* **c** *have $\ell_1$ distance close to one.*

PROOF: (Sketch) Let $A_S$ be the distribution that is uniform over $S \subseteq \{1, \ldots, n\}$. Pick $S$ at random among sets of size $n/2$ and run the algorithm on $A_S$. The algorithm only learns $o(n)$ elements from $S$. So with high probability the $\ell_1$ distance of whatever distribution the algorithm output will have $\ell_1$ distance from $A_S$ of nearly one. □

## 4 Applications to Markov Chains

Random walks on Markov chains generate probability distributions over the states of the chain, induced by the endpoints of the random walks. We employ $\ell_1$-**Distance-Test**, described in Section 2, to test mixing properties of Markov Chains.

This application of $\ell_1$-**Distance-Test** is initially inspired by the work of Goldreich and Ron [32], which conjectured an algorithm for testing expansion of bounded-degree graphs. Their algorithm is based on comparing the distribution of the endpoints of random walks on a graph to the uniform distribution via collisions. Subsequently to this work, Czumaj and Sohler [22], Kale and Seshadri [37], and Nachmias and Shapira [43] have independently concluded that the algorithm of Goldreich and Ron is provably a test for expansion property of graphs.

### 4.1 Preliminaries and Notation

Let **M** be a Markov chain represented by the transition probability matrix **M**. The point distribution $u$th state of **M** corresponds to an $n$-vector $\mathbf{e}_u = (0, \ldots, 1, \ldots, 0)$, with a one in only the $u$th location and zeroes elsewhere. The distribution generated by $t$-step random walks starting at state $u$ is denoted as a vector-matrix product $\mathbf{e}_u \mathbf{M}^t$.

Instead of computing such products in our algorithms, we assume that our $\ell_1$-**Distance-Test** has access to an oracle, `next_node` which on input of the state $u$ responds with the state $v$ with probability $\mathbf{M}(u, v)$. Given such an oracle, the distribution $\mathbf{e}_u^T \mathbf{M}^t$ can be generated in $O(t)$ steps.

Furthermore, the oracle itself can be realized in $O(\log n)$ time per query, given linear preprocessing time to compute the cumulative sums $\mathbf{M}_c(j,k) = \sum_{i=1}^{k} \mathbf{M}(j,i)$. The oracle can be simulated on input $u$ by producing a random number $\alpha$ in $[0,1]$ and performing binary search over the $u$th row of $\mathbf{M}_c$ to find $v$ such that $\mathbf{M}_c(u,v) \leq \alpha \leq \mathbf{M}_c(u, v+1)$. It then outputs state $v$. Note that when $\mathbf{M}$ is such that every row has at most $d$ nonzero terms, slight modifications of this yield an $O(\log d)$ implementation consuming $O(n+m)$ words of memory if $\mathbf{M}$ is $n \times n$ and has $m$ nonzero entries. Improvements of the work given in [55] can be used to prove that in fact constant query time is achievable with space consumption $O(n+m)$ for implementing `next_node`, given linear preprocessing time.

We define a notion of closeness between states $u$ and $v$, based on the distributions of endpoints of $t$ step random walks starting at $u$ and $v$ respectively.

**Definition 24** *We say that two states $u$ and $v$ are $(\epsilon, t)$-close if the distribution generated by $t$-step random walks starting at $u$ and $v$ are within $\epsilon$ in the $L_1$ norm, i.e. $\|\mathbf{e}_u \mathbf{M}^t - \mathbf{e}_v \mathbf{M}^t\|_1 < \epsilon$. Similarly we say that a state $u$ and a distribution $\mathbf{s}$ are $(\epsilon, t)$-close if $\|\mathbf{e}_u \mathbf{M}^t - \mathbf{s}\|_1 < \epsilon$.*

We say $\mathbf{M}$ is $(\epsilon, t)$-*mixing* if all states are $(\epsilon, t)$-close to the same distribution:

**Definition 25** *A Markov chain $\mathbf{M}$ is $(\epsilon, t)$-mixing if a distribution $\mathbf{s}$ exists such that for all states $u$, $\|\mathbf{e}_u \mathbf{M}^t - \mathbf{s}\|_1 \leq \epsilon$.*

For example, if $\mathbf{M}$ is $(\epsilon, O(\log n \log 1/\epsilon))$-mixing, then $\mathbf{M}$ is *rapidly-mixing* [53]. It can be easily seen that if $\mathbf{M}$ is $(\epsilon, t_0)$-mixing then it is $(\epsilon, t)$ mixing for all $t > t_0$.

We now make the following definition:

**Definition 26** *The* average $t$-step distribution, $\mathbf{s}_{\mathbf{M},t}$ *of a Markov chain $\mathbf{M}$ with $n$ states is the distribution*
$$\mathbf{s}_{\mathbf{M},t} = \frac{1}{n} \sum_u \mathbf{e}_u \mathbf{M}^t.$$

This distribution can be easily generated by picking $u$ uniformly from $[n]$ and walking $t$ steps from state $u$. In an $(\epsilon, t)$-mixing Markov chain, the average $t$-step distribution is $\epsilon$-close to the stationary distribution. In a Markov chain that is not $(\epsilon, t)$-mixing, this is not necessarily the case.

Each test given below assumes access to an $\ell_1$ distance tester $\ell_1$-**Distance-Test**$(u, v, \epsilon, \delta)$ which given oracle access to distributions $\mathbf{e}_u, \mathbf{e}_v$ over the same $n$ element set decides whether $\|\mathbf{e}_u - \mathbf{e}_v\|_1 \leq f(\epsilon)$ or if $\|\mathbf{e}_u - \mathbf{e}_v\|_1 > \epsilon$ with confidence $1 - \delta$. The time complexity of $L_1$_test is $T(n, \epsilon, \delta)$, and $f$ is the *gap* of the tester. The implementation of $\ell_1$-**Distance-Test** given earlier in Section 2 has gap $f(\epsilon) = \epsilon/(4\sqrt{n})$, and time complexity $T = O(\epsilon^{-8/3} n^{2/3} \log \frac{n}{\delta})$.

## 4.2 A Test for Mixing and a Test for Almost-Mixing

We show how to decide if a Markov chain is $(\epsilon, t)$-mixing; then, we define and solve a natural relaxation of that problem.

In order to test whether $\mathbf{M}$ is $(\epsilon, t)$-mixing, one can use $\ell_1$-**Distance-Test** to compare each distribution $\mathbf{e}_u \mathbf{M}^t$ with $\mathbf{s}_{\mathbf{M},t}$, with error parameter $\epsilon$ and confidence $\delta/n$. The running time is $O(nt \cdot T(n, \epsilon, \delta/n))$. The algorithm is given in Figure 4.

The behavior of the test is as follows: If every state is $(f(\epsilon)/2, t)$-close to some distribution $\mathbf{s}$, then $\mathbf{s}_{\mathbf{M},t}$ is $f(\epsilon)/2$-close to $\mathbf{s}$. Therefore every state is $(\epsilon, t)$-close to $\mathbf{s}_{\mathbf{M},t}$ and the tester passes. On

**Mixing**($\mathbf{M}, t, \epsilon, \delta$)

1. For each state $u$ in $\mathbf{M}$
   Reject if $\ell_1$-**Distance-Test**($\mathbf{e}_u \mathbf{M}^t, \mathbf{s}_{\mathbf{M},t}, \epsilon, \delta/n$) rejects.

2. Otherwise, accept.

Figure 4: Algorithm Mixing

the other hand, if there is no distribution that is $(\epsilon, t)$-close to all states, then, in particular, $\mathbf{s}_{\mathbf{M},t}$ is not $(\epsilon, t)$-close to at least one state and so the tester fails. Thus, we have shown the following theorem.

**Theorem 27** *Let $\mathbf{M}$ be a Markov chain. Given $\ell_1$-**Distance-Test** with time complexity $T(n, \epsilon, \delta)$ and gap $f$ and an oracle for* `next_node`*, there exists a test with time complexity $O(nt \cdot T(n, \epsilon, \delta/n))$ with the following behavior: If $\mathbf{M}$ is $(f(\epsilon)/2, t)$-mixing then $\Pr[\mathbf{M}$ is accepted$] > 1 - \delta$; if $\mathbf{M}$ is not $(\epsilon, t)$-mixing then $\Pr[\mathbf{M}$ is accepted$] < \delta$.*

For the implementation of $\ell_1$-**Distance-Test** given in Section 2, the running time of **Mixing** algorithm is $O(\epsilon^{-8/3} n^{5/3} t \log \frac{n}{\delta})$. It distinguishes between chains which are $\epsilon/(4\sqrt{n})$ mixing and those which are not $\epsilon$-mixing. The running time is sublinear in the size of $\mathbf{M}$ if $t \in o(n^{1/3}/\log(n))$.

A relaxation of this procedure is testing that *most* starting states reach the same distribution after $t$ steps. If $(1 - \rho)$ fraction of the states $u$ of a given $\mathbf{M}$ satisfy $\|\vec{s} - \mathbf{e}_u \mathbf{M}^t\|_1 \leq \epsilon$, then we say that $\mathbf{M}$ is $(\rho, \epsilon, t)$-*almost mixing*. The algorithm in Figure 5 tests whether a Markov chain is $(\rho, \epsilon, t)$-almost mixing.

**AlmostMixing**($\mathbf{M}, t, \epsilon, \delta, \rho$)
Repeat $O(1/\rho \cdot \ln(1/\delta))$ times

1. Pick a state $u$ in $\mathbf{M}$ uniformly at random.

2. Reject if $\ell_1$-**Distance-Test**($\mathbf{e}_u \mathbf{M}^t, \mathbf{s}_{\mathbf{M},t}, \epsilon, \delta\rho$) rejects.

Accept if none of the tests above rejected.

Figure 5: Algorithm AlmostMixing

Thus, we obtain the following theorem.

**Theorem 28** *Let $\mathbf{M}$ be a Markov chain. Given $\ell_1$-**Distance-Test** with time complexity $T(n, \epsilon, \delta)$ and gap $f$ and an oracle for* `next_node`*, there exists a test with time complexity $O(\frac{t}{\rho} T(n, \epsilon, \delta\rho) \log \frac{1}{\delta})$ with the following behavior: If $\mathbf{M}$ is $(\rho, f(\epsilon)/2, t)$-almost mixing then $\Pr[\mathbf{M}$ is accepted$] > 1 - \delta$; If $\mathbf{M}$ is not $(\rho, \epsilon, t)$-almost mixing then $\Pr[\mathbf{M}$ is accepted$] < \delta$.*

### 4.3 A Property Tester for Mixing

The main result of this section is a test that determines if a Markov chain's matrix representation can be changed in an $\epsilon$ fraction of the non-zero entries to turn it into a $(4\epsilon, 2t)$-mixing Markov chain. This notion falls within the scope of property testing [50, 30, 31, 23, 47], which in general takes a set $S$ with distance function $\Delta$ and a subset $P \subseteq S$ and decides if an elements $x \in S$ is in $P$ or if it is far from every element in $P$, according to $\Delta$. For the Markov chain problem, we take as our set $S$ all matrices $\mathbf{M}$ of size $n \times n$ with at most $d$ non-zero entries in each row. The distance function is given by the fraction of non-zero entries in which two matrices differ, and the difference in their average $t$-step distributions.

### 4.3.1 Preliminaries

We start with defining a distance function on a pair of Markov chains on the same state space.

**Definition 29** *Let $\mathbf{M}_1$ and $\mathbf{M}_2$ be $n$-state Markov chains with at most $d$ non-zero entries in each row. Define distance function $\Delta(\mathbf{M}_1, \mathbf{M}_2) = (\epsilon_1, \epsilon_2)$ if and only if $\mathbf{M}_1$ and $\mathbf{M}_2$ differ on $\epsilon_1 dn$ entries and $\|\mathbf{s}_{\mathbf{M}_1,t} - \mathbf{s}_{\mathbf{M}_2,t}\|_1 = \epsilon_2$. We say that $\mathbf{M}_1$ and $\mathbf{M}_2$ are $(\epsilon_1, \epsilon_2)$-close if $\Delta(\mathbf{M}_1, \mathbf{M}_2) \leq (\epsilon_1, \epsilon_2)$.*[2]

A natural question is whether all Markov chains are $\epsilon$-close to an $(\epsilon, t)$-mixing Markov chain, for certain parameters of $\epsilon$. For example, given a strongly connected and dense enough Markov chain, adding the edges of a constant-degree expander graph and choosing $t = \Theta(\log n)$ yields a Markov chain which $(\epsilon, t)$-mixes. However, for sparse Markov chains or small $\epsilon$, such a transformation does not work. Furthermore, the situation changes when asking whether there is an $(\epsilon, t)$-mixing Markov chain that is close both in the matrix representation and in the average $t$-step distribution: specifically, it can be shown that there exist constants $\epsilon, \epsilon_1, \epsilon_2 < 1$ and Markov chain $\mathbf{M}$ for which no Markov chain is both $(\epsilon_1, \epsilon_2)$-close to $\mathbf{M}$ and $(\epsilon, \log n)$-mixing. In fact, when $\epsilon_1$ is small enough, the problem becomes nontrivial even for $\epsilon_2 = 1$. The Markov chain corresponding to random walks on the $n$-cycle provides an example which is not $(t^{-1/2}, 1)$-close to any $(\epsilon, t)$-mixing Markov chain.

**Overview** As before, our algorithm proceeds by taking random walks on the Markov chain and comparing final distributions by using the $\ell_1$-**Distance-Test**. We define three types of states. First, a *normal* state is one from which a random walk arrives at nearly the average $t$-step distribution. In the discussion which follows, $t$ and $\epsilon$ denote constant parameters fixed as input to the algorithm.

**Definition 30** *Given a Markov Chain $\mathbf{M}$, a state $u$ of the chain is* normal *if it is $(\epsilon, t)$-close to $\mathbf{s}_{\mathbf{M},t}$. That is if $\|\mathbf{e}_u \mathbf{M}^t - \mathbf{s}_{\mathbf{M},t}\|_1 \leq \epsilon$. A state is* bad *if it is not normal.*

Testing normality requires time $O(t \cdot T(n, \epsilon, \delta))$. Using this definition, the first two algorithms given in this section can be described as testing whether all (*resp.* most) states in $\mathbf{M}$ are *normal*. Additionally, we need to distinguish states which not only produce random walks which arrive near $\mathbf{s}_{\mathbf{M},t}$ but which have low probability of visiting a bad state. We call such states *smooth* states.

**Definition 31** *A state $\mathbf{e}_u$ in a Markov chain $\mathbf{M}$ is* smooth *if (a) $u$ is $(\epsilon, \tau)$-close to $\mathbf{s}_{\mathbf{M},t}$ for $\tau = t, \ldots, 2t$ and (b) the probability that a $2t$-step random walk starting at $\mathbf{e}_u$ visits a bad state is at most $\epsilon$.*

Testing smoothness of a state requires $O(t^2 \cdot T(n, \epsilon, \delta))$ time. Our property test merely verifies by random sampling that most states are smooth.

---
[2]We say $(x, y) \leq (a, b)$ if $x \leq a$ and $y \leq b$.

### 4.3.2 The Test

We present below algorithm **TestMixing** in Figure 6, which on input Markov chain **M** and parameter $\epsilon$ determines whether at least $(1 - \epsilon)$ fraction of the states of **M** are smooth according to two distributions: uniform and the average $t$-step distribution. Assuming access to $\ell_1$-**Distance-Test** with complexity $T(n, \epsilon, \delta)$, this test runs in time $O(\epsilon^{-2} t^2 T(n, \epsilon, \frac{1}{6t}))$.

**TestMixing(M, $t, \epsilon$)**

1. Let $k = \Theta(1/\epsilon)$.
2. Choose $k$ states $u_1, \ldots, u_k$ uniformly at random.
3. Choose $k$ states $u_{k+1}, \ldots, u_{2k}$ independently according to $\mathbf{s}_{\mathbf{M},t}$.
4. For $i = 1$ to $2k$
   (a) $u = \vec{e}_{u_i}$.
   (b) For $w = 1$ to $O(1/\epsilon)$ and $j = 1$ to $2t$
       i. $u = \text{next\_node}(\mathbf{M}, u)$
       ii. $\ell_1$-**Distance-Test**$(\mathbf{e}_u \mathbf{M}^t, \mathbf{s}_{\mathbf{M},t}, \epsilon, \frac{1}{6t})$
   (c) For $\tau = t$ to $2t$, $\ell_1$-**Distance-Test**$(\vec{e}_{u_i} \mathbf{M}^\tau, \mathbf{s}_{\mathbf{M},t}, \epsilon, \frac{1}{3t})$
5. Pass if all tests pass.

Figure 6: Algorithm TestMixing

The main lemma of this section says that any Markov chain that is accepted by our test is $(2\epsilon, 1.01\epsilon)$-close to a $(4\epsilon, 2t)$-mixing Markov chain. First, we describe the modification of $M$ that we later show is $(4\epsilon, 2t)$-mixing.

**Definition 32** *$F$ is a function from $n \times n$ matrices to $n \times n$ matrices such that $F(\mathbf{M})$ returns $\widetilde{\mathbf{M}}$ by modifying the rows corresponding to bad states of $\mathbf{M}$ to $\mathbf{e}_u$, where $u$ is any smooth state.*

An important feature of the transformation $F$ is that it does not affect the distribution of random walks originating from smooth states very much.

**Lemma 33** *Given a Markov chain $\mathbf{M}$ and any state $u \in M$ which is smooth. If $\widetilde{\mathbf{M}} = F(\mathbf{M})$, then, for any time $t \leq \tau \leq 2t$, $\|\mathbf{e}_u \mathbf{M}^\tau - \mathbf{e}_u \widetilde{\mathbf{M}}^\tau\|_1 \leq \epsilon$ and $\|\mathbf{s}_{\mathbf{M},t} - \mathbf{e}_u \widetilde{\mathbf{M}}^\tau\|_1 \leq 2\epsilon$.*

PROOF: Define $\Gamma$ as the set of all walks of length $\tau$ from $u$ in **M**. Partition $\Gamma$ into $\Gamma_B$ and $\bar{\Gamma}_B$ where $\Gamma_B$ is the subset of walks which visit a bad state. Let $\chi_{w,i}$ be an indicator function which equals 1 if walk $w$ ends at state $i$, and 0 otherwise. Let weight function $W(w)$ be defined as the probability that walk $w$ occurs. Finally, define the primed counterparts $\Gamma'$, etc. for the Markov chain $\widetilde{\mathbf{M}}$. Now the $i$th element of $\mathbf{e}_u \mathbf{M}^\tau$ is $\sum_{w \in \Gamma_B} \chi_{w,i} \cdot W(w) + \sum_{w \in \bar{\Gamma}_B} \chi_{w,i} \cdot W(w)$. A similar expression can be written for each element of $\mathbf{e}_u \widetilde{\mathbf{M}}^\tau$. Since $W(w) = W'(w)$ whenever $w \in \bar{\Gamma}_B$ it follows that $\|\mathbf{e}_u \mathbf{M}^\tau - \mathbf{e}_u \widetilde{\mathbf{M}}^\tau\|_1 \leq \sum_i \sum_{w \in \Gamma_B} \chi_{w,i} |W(w) - W'(w)| \leq \sum_i \sum_{w \in \Gamma_B} \chi_{w,i} W(w) \leq \epsilon$.

Additionally, since $\|\mathbf{s}_{\mathbf{M},t} - \mathbf{e}_u \mathbf{M}^\tau\|_1 \leq \epsilon$ by the definition of smooth, it follows that $\|\mathbf{s}_{\mathbf{M},t} - \mathbf{e}_u \widetilde{\mathbf{M}}^\tau\|_1 \leq \|\mathbf{s}_{\mathbf{M},t} - \mathbf{e}_u \mathbf{M}^\tau\|_1 + \|\mathbf{e}_u \mathbf{M}^\tau - \mathbf{e}_u \widetilde{\mathbf{M}}^\tau\|_1 \leq 2\epsilon$. □

We can now prove the main lemma.

**Lemma 34** *If according to both the uniform distribution and the distribution* $\mathbf{s}_{\mathbf{M},t}$, $(1-\epsilon)$ *fraction of the states of a Markov chain* $\mathbf{M}$ *are smooth, then the matrix* $\mathbf{M}$ *is* $(2\epsilon, 1.01\epsilon)$-*close to a matrix* $\widetilde{\mathbf{M}}$ *which is* $(4\epsilon, 2t)$-*mixing.*

PROOF: Let $\widetilde{\mathbf{M}} = F(\mathbf{M})$. $\widetilde{\mathbf{M}}$ and $\mathbf{M}$ differ on at most $\epsilon n(d+1)$ entries. This gives the first part of our distance bound. For the second we analyze $\|\mathbf{s}_{\mathbf{M},t} - \mathbf{s}_{\widetilde{\mathbf{M}},t}\|_1 = \frac{1}{n}\sum_u \|\mathbf{e}_u\mathbf{M}^t - \mathbf{e}_u\widetilde{\mathbf{M}}^t\|_1$ as follows. The sum is split into two parts, over the nodes which are smooth and those nodes which are not. For each of the smooth nodes $u$, Lemma 33 says that $\|\mathbf{e}_u\mathbf{M}^t - \mathbf{e}_u\widetilde{\mathbf{M}}^t\|_1 \leq \epsilon$. Nodes which are not smooth account for at most $\epsilon$ fraction of the nodes in the sum, and thus can contribute no more than $\epsilon$ absolute weight to the distribution $\mathbf{s}_{\widetilde{\mathbf{M}},t}$. The sum can be bounded now by $\|\mathbf{s}_{\mathbf{M},t} - \mathbf{s}_{\widetilde{\mathbf{M}},t}\|_1 \leq \frac{1}{n}((1-\epsilon)n\epsilon + \epsilon n) \leq 2\epsilon$.

In order to show that $\widetilde{\mathbf{M}}$ is $(4\epsilon, 2t)$-mixing, we prove that for every state $u$, $\|\mathbf{s}_{\mathbf{M},t} - \mathbf{e}_u\mathbf{M}^{2t}\|_1 \leq 4\epsilon$. The proof considers three cases: $u$ smooth, $u$ bad, and $u$ normal. The last case is the most involved.

If $u$ is smooth in the Markov chain $\mathbf{M}$, then Lemma 33 immediately tells us that $\|\mathbf{s}_{\mathbf{M},t} - \mathbf{e}_u\widetilde{\mathbf{M}}^{2t}\|_1 \leq 2\epsilon$. Similarly if $u$ is bad in the Markov chain $\mathbf{M}$, then in the chain $\widetilde{\mathbf{M}}$ any path starting at $u$ transitions to a smooth state $v$ in one step. Since $\|\mathbf{s}_{\mathbf{M},t} - \mathbf{e}_v\widetilde{\mathbf{M}}^{2t-1}\|_1 \leq 2\epsilon$ by Lemma 33, the desired bound follows.

If $\mathbf{e}_u$ is a normal state which is not smooth, then we need a more involved analysis of the distribution $\mathbf{e}_u\widetilde{\mathbf{M}}^{2t}$. We divide $\Gamma$, the set of all $2t$-step walks in $\mathbf{M}$ starting at $u$, into three sets, which we consider separately.

For the first set take $\Gamma_B \subseteq \Gamma$ to be the set of walks which visit a bad node before time $t$. Let $\mathbf{d}_b$ be the distribution over endpoints of these walks, that is, let $\mathbf{d}_b$ assign to state $i$ the probability that any walk $w \in \Gamma_B$ ends at state $i$. Let $w \in \Gamma_B$ be any such walk. If $w$ visits a bad state at time $\tau < t$, then in the new Markov chain $\widetilde{\mathbf{M}}$, $w$ visits a smooth state $v$ at time $\tau + 1$. Another application of Lemma 33 implies that $\|\mathbf{e}_v\widetilde{\mathbf{M}}^{2t-\tau-1} - \mathbf{s}_{\mathbf{M},t}\|_1 \leq 2\epsilon$. Since this is true for all walks $w \in \Gamma_B$, we find $\|\mathbf{d}_b - \mathbf{s}_{\mathbf{M},t}\|_1 \leq 2\epsilon$.

For the second set, let $\Gamma_S \subseteq \Gamma \setminus \Gamma_B$ be the set of walks not in $\Gamma_B$ which visit a smooth state at time $t$. Let $\mathbf{d}_s$ be the distribution over endpoints of these walks. Any walk $w \in \Gamma_S$ is identical in the chains $\mathbf{M}$ and $\widetilde{\mathbf{M}}$ up to time $t$, and then in the chain $\widetilde{\mathbf{M}}$ visits a smooth state $v$ at time $t$. Thus since $\|\mathbf{e}_v\widetilde{\mathbf{M}}^t - \mathbf{s}_{\mathbf{M},t}\|_1 \leq 2\epsilon$, we have $\|\mathbf{d}_s - \mathbf{s}_{\mathbf{M},t}\|_1 \leq 2\epsilon$.

Finally, let $\Gamma_N = \Gamma \setminus (\Gamma_B \cup \Gamma_S)$, and let $\mathbf{d}_n$ be the distribution over endpoints of walks in $\Gamma_N$. $\Gamma_N$ consists of a subset of the walks from a normal node $u$ which do not visit a smooth node at time $t$. By the definition of normal, $u$ is $(\epsilon, t)$-close to $\mathbf{s}_{\mathbf{M},t}$ in the Markov chain $\mathbf{M}$. By assumption at most $\epsilon$ weight of $\mathbf{s}_{\mathbf{M},t}$ is assigned to nodes which are not smooth. Therefore $|\Gamma_N|/|\Gamma|$ is at most $\epsilon + \epsilon = 2\epsilon$.

Now define the weights of these distributions as $\omega_b, \omega_s$ and $\omega_n$. That is $\omega_b$ is the probability that a walk from $u$ in $\mathbf{M}$ visits a bad state before time $t$. Similarly $\omega_s$ is the probability that a walk does not visit a bad state before time $t$, but visits a smooth state at time $t$, and $\omega_n$ is the probability that a walk does not visit a bad state but visits a normal, non-smooth state at time $t$. Then, $\omega_b + \omega_s + \omega_n = 1$. Finally, $\|\mathbf{e}_u\widetilde{\mathbf{M}}^{2t} - \mathbf{s}_{\mathbf{M},t}\|_1 = \|\omega_b\mathbf{d}_b + \omega_s\mathbf{d}_s + \omega_n\mathbf{d}_n - \mathbf{s}_{\mathbf{M},t}\|_1 \leq \omega_b\|\mathbf{d}_b - \mathbf{s}_{\mathbf{M},t}\|_1 + \omega_s\|\mathbf{d}_s - \mathbf{s}_{\mathbf{M},t}\|_1 + \omega_n\|\mathbf{d}_n - \mathbf{s}_{\mathbf{M},t}\|_1 \leq (\omega_b + \omega_s)\max\{\|\mathbf{d}_b - \mathbf{s}_{\mathbf{M},t}\|_1, \|\mathbf{d}_s - \mathbf{s}_{\mathbf{M},t}\|_1\} + \omega_n\|\mathbf{d}_n - \mathbf{s}_{\mathbf{M},t}\|_1 \leq 4\epsilon$. □

Given this, we finally can show our main theorem.

**Theorem 35** *Let $\mathbf{M}$ be a Markov chain. Given $\ell_1$-**Distance-Test** with time complexity $T(n, \epsilon, \delta)$ and gap $f$ and an oracle for* `next_node`*, there exists a test such that if $\mathbf{M}$ is $(f(\epsilon), t)$-mixing then the test accepts with probability at least $2/3$. If $\mathbf{M}$ is not $(2\epsilon, 1.01\epsilon)$-close to any $\widetilde{\mathbf{M}}$ which is $(4\epsilon, 2t)$-mixing then the test rejects with probability at least $2/3$. The runtime of the test is $O(\frac{1}{\epsilon^2} \cdot t^2 \cdot T(n, \epsilon, \frac{1}{6t}))$.*

PROOF: Since in any Markov chain $\mathbf{M}$ which is $(\epsilon, t)$-mixing all states are smooth, $\mathbf{M}$ accepts this test with probability at least $(1-\delta)$. Furthermore, any Markov chain with at least $(1-\epsilon)$ fraction of smooth states is $(2\epsilon, 1.01\epsilon)$-close to a Markov chain which is $(4\epsilon, 2t)$-mixing, by Lemma 34. □

### 4.4 Extension to Sparse Graphs and Uniform Distributions

The property test can also be made to work for general sparse Markov chains by a simple modification to the testing algorithms. Consider Markov chains with at most $m \ll n^2$ nonzero entries, but with no nontrivial bound on the number of nonzero entries per row. Then, the definition of the distance should be modified to $\Delta(M_1, M_2) = (\epsilon_1, \epsilon_2)$ if $M_1$ and $M_2$ differ on $\epsilon_1 \cdot m$ entries and the $\|\mathbf{s}_{\mathbf{M}_1,t} - \mathbf{s}_{\mathbf{M}_2,t}\|_1 = \epsilon_2$. The above test does not suffice for testing that $\mathbf{M}$ is $(\epsilon_1, \epsilon_2)$-close to an $(\epsilon, t)$-mixing Markov chain $\widetilde{M}$, since in our proof, the rows corresponding to bad states may have many nonzero entries and thus $\mathbf{M}$ and $\widetilde{M}$ may differ in a large fraction of the nonzero entries. However, let $D$ be a distribution on states in which the probability of each state is proportional to cardinality of the support set of its row. Natural ways of encoding this Markov chain allow constant time generation of states according to $D$. By modifying the algorithm to also test whether most states according to $D$ are smooth, one can show that $\mathbf{M}$ is close to an $(\epsilon, t)$-mixing Markov chain $\widetilde{M}$.

Because of our ability to test $\epsilon$-closeness to the *uniform* distribution in $O(n^{1/2}\epsilon^{-2})$ steps [32], it is possible to speed up our test for mixing for those Markov chains known to have uniform stationary distribution, such as Markov chains corresponding to random walks on regular graphs. An ergodic random walk on the vertices of an undirected graph instead may be regarded (by looking at it "at times $t + 1/2$") as a random walk on the *edge-midpoints* of that graph. The stationary distribution on edge-midpoints always exists and is uniform. So, for undirected graphs we can speed up mixing testing by using a tester for closeness to the uniform distribution.

**Acknowledgments** We are very grateful to Oded Goldreich and Dana Ron for sharing an early draft of their work with us and for several helpful discussions. We would also like to thank Naoke Abe, Richard Beigel, Yoav Freund, Russell Impagliazzo, Jeff Ketchersid, Kevin Matulef, Alexis Maciel, Krzysztof Onak, Sofya Raskhodnikova, and Tassos Viglas for helpful discussions. Finally, we thank Ning Xie for pointing out errors in the proofs in an earlier version.

# References

[1] Michal Adamaszek, Artur Czumaj, and Christian Sohler. Testing monotone continuous distributions on high-dimensional real cubes. In *Proceedings of 21st ACM-SIAM Symposium on Discrete Algorithms*, pages 56–65, 2010.

[2] N. Alon, M. Krivelevich, E. Fischer, and M. Szegedy. Efficient testing of large graphs. In IEEE, editor, *40th Annual Symposium on Foundations of Computer Science: October 17–19,*

1999, New York City, New York,, pages 656–666, 1109 Spring Street, Suite 300, Silver Spring, MD 20910, USA, 1999. IEEE Computer Society Press.

[3] N. Alon, Y. Matias, and M. Szegedy. The space complexity of approximating the frequency moments. *JCSS*, 58, 1999.

[4] Noga Alon. Eigenvalues and expanders. *Combinatorica*, 6(2):83–96, 1986.

[5] Noga Alon, Alexandr Andoni, Tali Kaufman, Kevin Matulef, Ronitt Rubinfeld, and Ning Xie. Testing k-wise and almost k-wise independence. In David S. Johnson and Uriel Feige, editors, *STOC*, pages 496–505. ACM, 2007.

[6] Ziv Bar-Yossef, Ravi Kumar, and D. Sivakumar. Sampling algorithms: Lower bounds and applications. In *Proceedings of 33th Symposium on Theory of Computing*, Crete, Greece, 6–8 July 2001. ACM.

[7] Tuğkan Batu, Sanjoy Dasgupta, Ravi Kumar, and Ronitt Rubinfeld. The complexity of approximating the entropy. *SIAM Journal on Computing*, 35(1):132–150, 2005.

[8] Tuğkan Batu, Lance Fortnow, Eldar Fischer, Ravi Kumar, Ronitt Rubinfeld, and Patrick White. Testing random variables for independence and identity. In *Proceedings of 42nd FOCS*. IEEE, 2001.

[9] Tuğkan Batu, Lance Fortnow, Ronitt Rubinfeld, Warren D. Smith, and Patrick White. Testing that distributions are close. In *Proceedings of the 41st Annual Symposium on Foundations of Computer Science*, pages 259–269, Redondo Beach, CA, 2000. IEEE Computer Society.

[10] Tuğkan Batu, Ravi Kumar, and Ronitt Rubinfeld. Sublinear algorithms for testing monotone and unimodal distributions. In *Proceedings of 36th ACM Symposium on Theory of Computing*, pages 381–390, 2004.

[11] Lakshminath Bhuvanagiri and Sumit Ganguly. Estimating entropy over data streams. In Yossi Azar and Thomas Erlebach, editors, *ESA*, volume 4168 of *Lecture Notes in Computer Science*, pages 148–159. Springer, 2006.

[12] Mickey Brautbar and Alex Samorodnitsky. Approximating entropy from sublinear samples. In Nikhil Bansal, Kirk Pruhs, and Clifford Stein, editors, *SODA*, pages 366–375. SIAM, 2007.

[13] Vladimir Braverman and Rafail Ostrovsky. Measuring independence of datasets. In *Proceedings of the 42nd ACM Symposium on Theory of Computing, STOC 2010, Cambridge, Massachusetts, USA, 5-8 June 2010*, pages 271–280, 2010.

[14] Vladimir Braverman and Rafail Ostrovsky. Zero-one frequency laws. In *Proceedings of the 42nd ACM Symposium on Theory of Computing, STOC 2010, Cambridge, Massachusetts, USA, 5-8 June 2010*, pages 281–290, 2010.

[15] A. Broder, M. Charikar, A. Frieze, and M. Mitzenmacher. Min-wise independent permutations. *JCSS*, 60, 2000.

[16] Amit Chakrabarti, Khanh Do Ba, and S. Muthukrishnan. Estimating entropy and entropy norm on data streams. In Bruno Durand and Wolfgang Thomas, editors, *STACS*, volume 3884 of *Lecture Notes in Computer Science*, pages 196–205. Springer, 2006.

[17] Amit Chakrabarti, Graham Cormode, and Andrew McGregor. A near-optimal algorithm for estimating the entropy of a stream. *ACM Transactions on Algorithms*, 6(3), 2010.

[18] Steve Chien, Katrina Ligett, and Andrew McGregor. Space-efficient estimation of robust statistics and distribution testing. In *Proceedings of Innovations in Computer Science*, Beijing, China, 2010.

[19] Thomas M. Cover and Joy A. Thomas. *Elements of Information Theory*. Wiley Series in Telecommunications. John Wiley & Sons, 1991.

[20] N. Cressie and P.B. Morgan. Design considerations for Neyman Pearson and Wald hypothesis testing. *Metrika*, 36(6):317–325, 1989.

[21] I. Csiszár. Information-type measures of difference of probability distributions and indirect observations. *Studia Scientiarum Mathematicarum Hungarica*, 1967.

[22] Artur Czumaj and Christian Sohler. Testing expansion in bounded-degree graphs. In *FOCS*, pages 570–578. IEEE Computer Society, 2007.

[23] Funda Ergün, Sampath Kannan, S. Ravi Kumar, Ronitt Rubinfeld, and Mahesh Viswanathan. Spot-checkers. In *STOC 30*, pages 259–268, 1998.

[24] J. Feigenbaum, S. Kannan, M. Strauss, and M. Viswanathan. An approximate $L^1$-difference algorithm for massive data streams (extended abstract). In *FOCS 40*, 1999.

[25] William Feller. *An Introduction to Probability Theory and Applications*, volume 1. John Wiley & Sons Publishers, New York, NY, 3rd ed., 1968.

[26] J. Fong and M. Strauss. An approximate $L^p$-difference algorithm for massive data streams. In *Annual Symposium on Theoretical Aspects of Computer Science*, 2000.

[27] Alan Frieze and Ravi Kannan. Quick approximation to matrices and applications. *COMBINAT: Combinatorica*, 19, 1999.

[28] Phillip B. Gibbons and Yossi Matias. Synopsis data structures for massive data sets. In *SODA 10*, pages 909–910. ACM-SIAM, 1999.

[29] O. Goldreich and L. Trevisan. Three theorems regarding testing graph properties. Technical Report ECCC-10, Electronic Colloquium on Computational Complexity, January 2001.

[30] Oded Goldreich, Shafi Goldwasser, and Dana Ron. Property testing and its connection to learning and approximation. In *FOCS 37*, pages 339–348. IEEE, 14–16 October 1996.

[31] Oded Goldreich and Dana Ron. Property testing in bounded degree graphs. In *STOC 29*, pages 406–415, 1997.

[32] Oded Goldreich and Dana Ron. On testing expansion in bounded-degree graphs. Technical Report TR00-020, Electronic Colloquium on Computational Complexity, 2000.

[33] G. H. Golub and C. F. van Loan. *Matrix Computations*. The John Hopkins University Press, Baltimore, MD, 1996.

[34] Sudipto Guha, Piotr Indyk, and Andrew McGregor. Sketching information divergences. *Machine Learning*, 72(1-2):5–19, 2008.

[35] Sudipto Guha, Andrew McGregor, and Suresh Venkatasubramanian. Sublinear estimation of entropy and information distances. *ACM Transactions on Algorithms*, 5(4), 2009.

[36] Piotr Indyk and Andrew McGregor. Declaring independence via the sketching of sketches. In *Proceedings of the Nineteenth Annual ACM-SIAM Symposium on Discrete Algorithms, SODA 2008, San Francisco, California, USA, January 20-22, 2008*, pages 737–745, 2008.

[37] Satyen Kale and C. Seshadhri. An expansion tester for bounded degree graphs. In Luca Aceto, Ivan Damgård, Leslie Ann Goldberg, Magnús M. Halldórsson, Anna Ingólfsdóttir, and Igor Walukiewicz, editors, *ICALP (1)*, volume 5125 of *Lecture Notes in Computer Science*, pages 527–538. Springer, 2008.

[38] R. Kannan. Markov chains and polynomial time algorithms. In Shafi Goldwasser, editor, *Proceedings: 35th Annual Symposium on Foundations of Computer Science, November 20–22, 1994, Santa Fe, New Mexico*, pages 656–671, 1109 Spring Street, Suite 300, Silver Spring, MD 20910, USA, 1994. IEEE Computer Society Press.

[39] Sampath Kannan and Andrew Chi-Chih Yao. Program checkers for probability generation. In Javier Leach Albert, Burkhard Monien, and Mario Rodríguez-Artalejo, editors, *ICALP 18*, volume 510 of *Lecture Notes in Computer Science*, pages 163–173, Madrid, Spain, 8–12 July 1991. Springer-Verlag.

[40] Donald E. Knuth. *The Art of Computer Programming, Volume III: Sorting and Searching*. Addison-Wesley, 1973.

[41] E. L. Lehmann. *Testing Statistical Hypotheses*. Wadsworth and Brooks/Cole, Pacific Grove, CA, second edition, 1986. [Formerly New York: Wiley].

[42] S-K. Ma. Calculation of entropy from data of motion. *Journal of Statistical Physics*, 26(2):221–240, 1981.

[43] Asaf Nachmias and Asaf Shapira. Testing the expansion of a graph. *Electronic Colloquium on Computational Complexity (ECCC)*, 14(118), 2007.

[44] J. Neyman and E.S. Pearson. On the problem of the most efficient test of statistical hypotheses. *Philos. Trans. Royal Soc. A*, 231:289–337, 1933.

[45] Liam Paninski. A coincidence-based test for uniformity given very sparsely sampled discrete data. *IEEE Transactions on Information Theory*, 54(10):4750–4755, 2008.

[46] Beresford N. Parlett. *The Symmetric Eigenvalue Problem*, volume 20 of *Classics in applied mathematics*. Society for Industrial and Applied Mathematics, Philadelphia, PA, USA, 1998.

[47] Michal Parnas and Dana Ron. Testing the diameter of graphs. In Dorit Hochbaum, Klaus Jensen, José D.P. Rolim, and Alistair Sinclair, editors, *Randomization, Approximation, and Combinatorial Optimization*, volume 1671 of *Lecture Notes in Computer Science*, pages 85–96. Springer-Verlag, 8–11 August 1999.

[48] Sofya Raskhodnikova, Dana Ron, Amir Shpilka, and Adam Smith. Strong lower bounds for approximating distribution support size and the distinct elements problem. *SIAM J. Comput.*, 39(3):813–842, 2009.

[49] Ronitt Rubinfeld and Rocco A. Servedio. Testing monotone high-dimensional distributions. *Random Struct. Algorithms*, 34(1):24–44, 2009.

[50] Ronitt Rubinfeld and Madhu Sudan. Robust characterizations of polynomials with applications to program testing. *SIAM Journal on Computing*, 25(2):252–271, April 1996.

[51] Ronitt Rubinfeld and Ning Xie. Testing non-uniform -wise independent distributions over product spaces. In Samson Abramsky, Cyril Gavoille, Claude Kirchner, Friedhelm Meyer auf der Heide, and Paul G. Spirakis, editors, *ICALP (1)*, volume 6198 of *Lecture Notes in Computer Science*, pages 565–581. Springer, 2010.

[52] Amit Sahai and Salil Vadhan. A complete promise problem for statistical zero-knowledge. In *Proceedings of the 38th Annual Symposium on the Foundations of Computer Science*, pages 448–457. IEEE, 20–22 October 1997.

[53] Alistair Sinclair and Mark Jerrum. Approximate counting, uniform generation and rapidly mixing Markov chains. *Information and Computation*, 82(1):93–133, July 1989.

[54] Paul Valiant. Testing symmetric properties of distributions. In *Proceedings of the 40th Annual ACM Symposium on Theory of Computing*, pages 383–392, 2008.

[55] A. J. Walker. An efficient method for generating discrete random variables with general distributions. *ACM trans. math. software*, 3:253–256, 1977.

[56] Kenji Yamanishi. Probably almost discriminative learning. *Machine Learning*, 18(1):23–50, 1995.

## A  Chebyshev's Inequality

Chebyshev's inequality states that for any random variable $A$, and $\rho > 0$,

$$\Pr\left[\,|A - E[A]| \ge \rho\,\right] \le \frac{\operatorname{Var}(A)}{\rho^2}.$$